\documentclass[preprint,a4paper,10pt,3p]{elsarticle}

\usepackage{amssymb}

\usepackage{lineno}

\newdefinition{rmk}{Remark}

\usepackage{physics}
\usepackage{xcolor}
\usepackage{psfrag}
\usepackage{setspace}
\usepackage{textcomp}

\journal{arxiv}

\begin{document}

\begin{frontmatter}


\title{What do cells regulate in soft tissues on short time scales?}

\author[1,2]{Jonas F. Eichinger\corref{cor}}
\ead{eichinger@lnm.mw.tum.de, phone: +49 (0) 89 289 15252}

\author[2,3]{Daniel Paukner}
\ead{daniel.paukner@hzg.de}

\author[3]{Roland C. Aydin}
\ead{roland.aydin@hzg.de}

\author[1]{Wolfgang A. Wall}
\ead{wall@lnm.mw.tum.de}

\author[4]{Jay D. Humphrey}
\ead{jay.humphrey@yale.edu}

\author[2,3]{Christian J. Cyron}
\ead{christian.cyron@tuhh.de}

\cortext[cor]{corresponding author}

\address[1]{Institute for Computational Mechanics, Technical University of Munich, Boltzmannstrasse 15, 85748, Garching, Germany}

\address[2]{Institute of Continuum and Materials Mechanics, Hamburg University of Technology, Eissendorfer Str. 42, 21073, Hamburg, Germany}

\address[3]{Institute of Material Systems Modeling, Helmholtz-Zentrum Geesthacht, Max-Planck-Strasse 1, 21502, Geesthacht, Germany}

\address[4]{Department of Biomedical Engineering, Yale University, 55 Prospect Street, New Haven, CT 06520 USA}

\begin{abstract}
Cells within living soft biological tissues seem to promote the maintenance of a mechanical state within a defined range near a so-called set-point. This mechanobiological process is often referred to as mechanical homeostasis. During this process, cells intimately interact with the fibers of the surrounding extracellular matrix (ECM). It remains poorly understood, however, what individual cells actually regulate during these interactions, and how these micromechanical regulations are translated to tissue level to lead to what we macroscopically call mechanical homeostasis. Herein, we examine this question by a combination of experiments, theoretical analysis and computational modeling. We demonstrate that on short time scales (hours) - during which deposition and degradation of ECM fibers can largely be neglected - cells appear to regulate neither the stress / strain in the ECM nor their own shape, but rather only the contractile forces that they exert on the surrounding ECM.
\end{abstract}



\begin{keyword}
homeostasis \sep mechanosensation \sep mechanoregulation \sep cell-matrix interactions \sep discrete fiber model
\end{keyword}

\end{frontmatter}

\section{Introduction} 
\label{sec:intro}

\renewcommand{\thepage}{\arabic{page}}
\setcounter{page}{1}

While many engineering materials remain stress-free, or in their respective production-induced eigenstress state, in the absence of external loading, living soft tissues generally seek to establish a preferred mechanical state that is not stress-free. This state is often referred to as \textit{homeostatic}. Albeit near steady state, cells are yet highly active. Cells constantly sense and transduce environmental cues into intracellular signaling pathways (mechanosentation) and  adjust their interactions with the surrounding tissue fibers (mechanoregulation) accordingly \cite{Tomasek2002,Humphrey2014,Jiang2006,Cavalcanti-Adam2007,Lerche2020}. To this end, cells use transmembrane receptors such as integrins to connect the intracellular cytoskeleton to fibers of the extracellular matrix (ECM). This unique dynamic regulatory system allows cells to establish and maintain a preferred mechanical state, which is often referred to as \textit{tensional} \cite{Brown1998} or \textit{mechanical} \cite{Humphrey2008} \textit{homeostasis}. It has been shown that compromised or lost mechanical homeostasis, and its underlying mechanosensitive and mechanoregulatory processes, are intimately linked to some of the most predominant causes of death, such as aneurysms \cite{Humphrey2014a} or cancer \cite{Weaver1997,Paszek2005,Yeung2005,Levental2009,Butcher2009,Lu2012} on the organ scale $[cm]$, and to cellular processes such as cell migration \cite{Kim2020,Xie2017,Hall2016,Grinnell2010}, differentiation \cite{Chiquet2009,Mammoto2012,Zemel2015}, and even survival \cite{Bates1995,ZHU2002,Sukharev2012,Schwartz1995}. 

Despite the prominent role of mechanical homeostasis in various physiological and pathophysiological processes, it remains unclear which mechanical quantity is regulated on a tissue level. In simple tissue equivalents, it has been hypothesized that this ubiquitous control may seek to develop and maintain a certain state of tension in the tissue. Although continuum metrics of stress, strain, and those derived from them are unlikely to be sensed directly by cells \cite{Humphrey2001}, such metrics can nevertheless be good surrogate markers sufficient for data analysis and computation \cite{Brown1998,Ezra2010,Eichinger2020,Wolinsky1967,Shadwick1999,Elosegui-Artola2016,Elosegui-Artola2018}. To address this open question, experiments using tissue equivalents have attracted increasing attention over the last decades \cite{Eichinger2020b}. Tissue equivalents are simple model systems of living soft tissues and consist often of collagen fibers seeded with living cells. When fixed at their boundaries in an initially stress-free configuration, tissue equivalents exhibit a characteristic behavior observed in numerous independent studies \cite{Brown1998,Ezra2010,Eichinger2020,Marenzana2006,Eichinger2020b,Brown2002,Courderot-Masuyer2017,Campbell2003a,Dahlmann-Noor2007,Karamichos2007,Sethi2002}. First, they rapidly build up a certain level of internal tension (phase I). Second, this level of tension is maintained for a prolonged period (phase II). If this steady state is perturbed (e.g., by stretching or releasing the tissue equivalent slightly), cells seems to regulate their activity such that the tension in the gel is restored towards the value prior to the perturbation \cite{Brown1998,Ezra2010,Eichinger2020}. It remains unclear, however, whether this value is recovered within a range consistent with homeostasis, noting that "homeo" means similar to in contrast with "homo" which means the same as \cite{Cannon1929}.

In general, different time scales are involved in mechanical homeostasis. On short time scales (minutes to hours), cells can adapt the forces they exert on the surrounding ECM. On longer time scales (several days to months), cells may additionally turnover the ECM, that is, inelastically reorganize its microstructure or deposit and degrade matrix fibers (growth and atrophy) \cite{Marenzana2006,Simon2014,Matsumoto1994,Nakagawa1989}.
This article focuses on short time scales, in which the regulation of cellular forces can be assumed to be the dominant mechanism of mechanical homeostasis. Not only different time scales, but also different spatial scales are involved. On the microscale, individual cells can probably sense and regulate elementary quantities such as forces in or displacements of surrounding fibers \cite{Humphrey2001}. By contrast, on the tissue scale, this cellular activity leads to changes of continuum mechanical quantities such as stress, strain, or stiffness. 

In this paper, we consider the question of which mechanical quantity individual cells regulate on the microscale on short time scales (where growth and remodeling can largely be neglected), and how this behavior translates into changes of continuum mechanical quantities on the tissue level. We address this question by a combination of three tools. First, we performed biaxial tissue culture experiments with a custom-built bioreactor \cite{Eichinger2020}. Second, we developed a simple theoretical mechanical analog model to understand the governing principles of our experimental observations. Third, we used a detailed computational model resolving cell-ECM interactions on the level of discrete cells and fibers \cite{Eichinger2020c} to validate the results of our theoretical analysis.

\section{Material and methods}
\label{sec:methods}

\subsection{Preparation of tissue equivalents for experiments}
\label{subsec:methods_exp}
Primary smooth muscle cells (SMCs) were isolated from 13--15 week old male C57BL/6 wild-type mouse aortas. Cells extracted from the medial layer of the descending, suprarenal, and infrarenal aorta (all having a mesoderm embryonic lineage \cite{Majesky2007}) were mixed and then expanded in culture. Cells were maintained in culture medium consisting of Dulbecco's Modified Eagles's Medium (DMEM) (Gibco, Life Technologies, D5796), $20\%$ heat-inactivated fetal bovine serum (FBS) (Gibco, Life Technologies), and $1\%$ penicillin-streptomycin (ThermoFisher) in an incubator at $37^\circ\text{C}$ and $5\%$ $CO_2$. After cell extraction, cells were grown in one well of a 6-well plate before being transferred to a T25 flask in passage 1 (P1). In P2 and P3, cells were grown in T75 flasks. Cells were passaged at $70-80\%$ confluence roughly every 6 days. Passages 4 and 5 were used in all experiments. Cells were starved in medium containing $2.0\%$ FBS for $24h$ prior to the experiments to inhibit proliferation during the experiments.

\begin{figure}[htb!] 
\begin{center}
\includegraphics[width=0.95\textwidth]{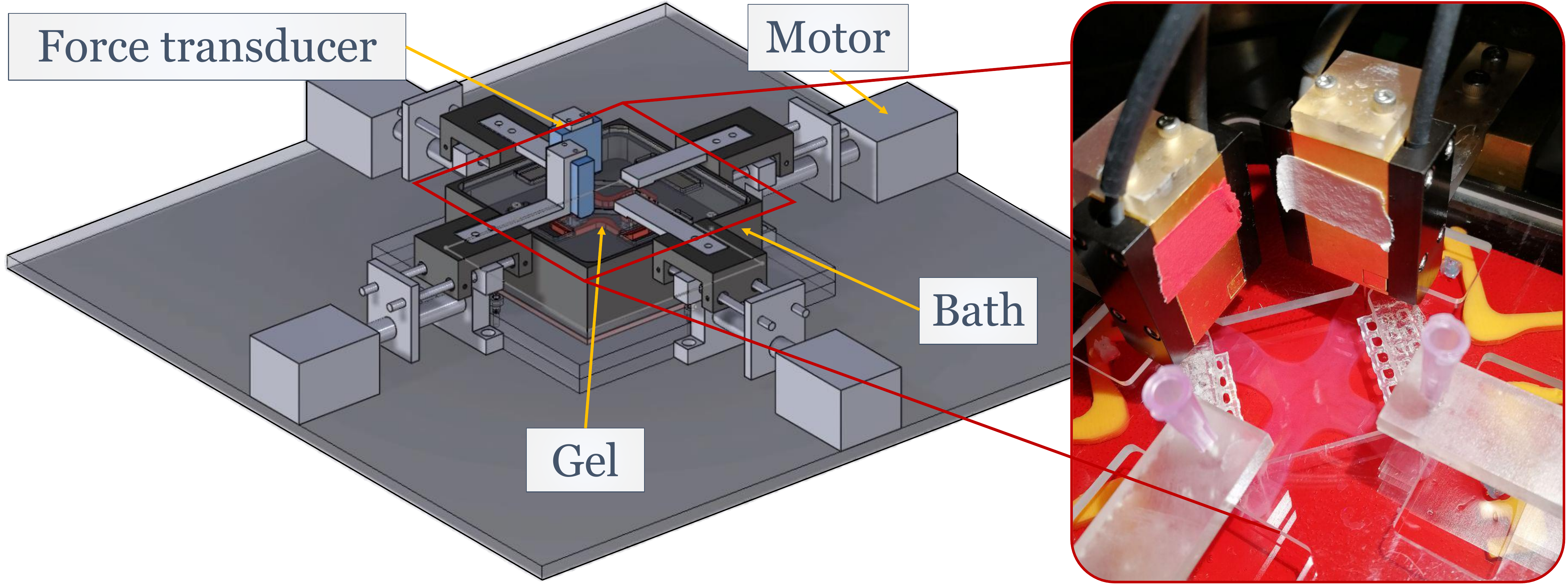}
\end{center}  
\caption{Biaxial testing device for cell-seeded collagen gels as introduced in \cite{Eichinger2020}: schematic (left) and experimental set-up (right).}
\label{fig:exp_device}       
\end{figure}

SMC-seeded collagen gels were prepared on ice following a protocol slightly modified from \cite{Eichinger2020}. Briefly, $1.428ml$ of $5x$ DMEM, $0.683ml$ of a $10x$ reconstitution buffer (0.1 N NaOH and 20 mM HEPES; Sigma), and $0.790ml$ of high concentration, type-I rat tail collagen ($8.22 mg/ml$; Corning) were mixed with $4.1ml$ of experimental culture medium containing $3.5 \cdot 10^{6}$ SMCs for a total volume of $7.0ml$ of gel solution. This resulted in a collagen concentration of 1.0 mg/ml and a cell density of $0.5 \times 10^6$ cells/ml. The experimental culture medium consisted of DMEM supplemented with $2.0\%$ FBS and $1\%$ penicillin-streptomycin. The final gel solution was pipetted into a cruciform mold placed within a custom-built biaxial bioreactor \cite{Eichinger2020} (Fig. \ref{fig:exp_device}). The mold was removed after 30 - 45 minutes of gelation, and the bath was filled with $80 ml$ of the experimental culture medium. This detached the gel from the base of the bath and allowed it to float freely. The initial stiffness of these gels in the small deformation regime, that is, their Young\textquotesingle s modulus, was measured to be about $1 kPa$.

\subsection{Mechanical analog model of soft tissue mechanical homeostasis on short time scales}
\label{subsec:analog_model}

\begin{figure*}[htb!] 
\begin{center}
 \includegraphics[width=0.75\textwidth]{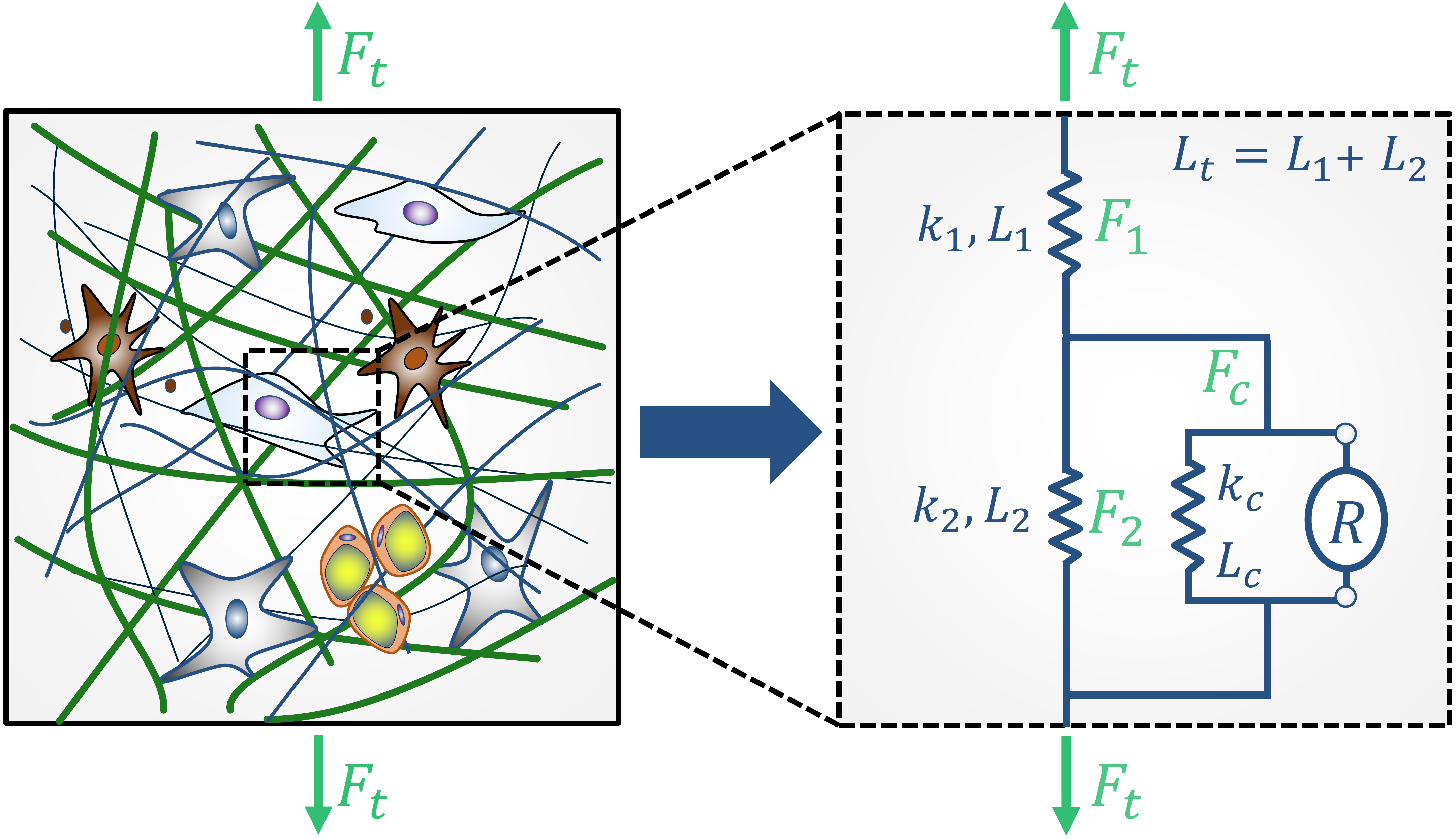} \end{center}  
\caption{Mechanical analog model of a three-dimensional fiber matrix with embedded cells: the force of all cellular forces in the direction of interest is given by $F_c$ composed of an active component mediated by the regulator element $R$ and a passive component. In this way, cells pull on ECM fibers. These fibers are connected via the network to other fibers parallel to the cell (region 2), which are in general compressed when the cell exerts contractile forces. Both sets of fibers are represented by elastic springs. The resulting force on tissue level (as measured, for example, by force sensors at clamped boundaries) corresponds to the force $F_t$ in the mechanical analog model. Note, the analog model can also be understood as the smallest possible RVE for soft tissues.}
\label{fig:intro_of_analog_model}       
\end{figure*}

To understand the underlying principles of mechanical homeostasis of soft tissues, we developed a simplified mechanical analog model (Fig. \ref{fig:intro_of_analog_model}). The mechanical environment of cells was modeled as an elastic network of fibers. Viscoelastic effects were neglected because in the ECM they manifest on time scales much shorter than that for mechanical homeostasis. In our mechanical analog model, cells are represented by an elastic spring (representing their passive stiffness) with a regulator element in parallel. The latter represents the forces exerted by the stress fibers in the cytoskeleton on the surrounding ECM fibers. By these forces, fibers connected to the cells via focal adhesions are stretched. At the same time, fibers aligned in the same direction but beside the cells are necessarily shortened as the cells contract. Note that this is true in any direction and can occur in several independent spatial directions at the same time in case of a multi-axial stress state. For simplicity, we focus only on a single direction, noting that an analogous discussion would be possible in any other direction. The above scenario in a single direction leads to the mechanical analog model depicted in Fig. \ref{fig:intro_of_analog_model}. In general, the mechanical function of the fibers is represented by elastic spring elements. The forces in the spring elements (i.e., forces transmitted through all the fibers of category 1 and 2 with unit [N], respectively) are denoted by $F_1$ and $F_2$. The force exerted by all cells in the direction of interest is denoted by $F_c$. It is composed of an active component exerted by the regulator element $R$ and a passive component. Generally, the passive elastic forces of the different elements in our mechanical analog model are characterized by the overall stiffness $k_i$ and a length $L_i$ in the direction of interest with $i \in \{1,2,c\}$. That is, for the passive elastic parts of our model, changes of length and force are related by
\begin{align}
\Delta F_i &= k_i \Delta L_i, \quad i \in \{1,2,c\}.
\label{eqn:force_ext}
\end{align}
The overall force of the tissue (with length $L_t$) in the direction of interest is denoted by $F_t$. It is important to note that this model can also be interpreted as the smallest possible representative volume element (RVE) of a uniaxially loaded or constrained soft tissue.

\subsection{Three-dimensional discrete fiber and cell model}
\label{subsec:discrete_fiber_model}

\begin{figure}[htb!] 
\begin{center}
\includegraphics[width=0.85\textwidth]{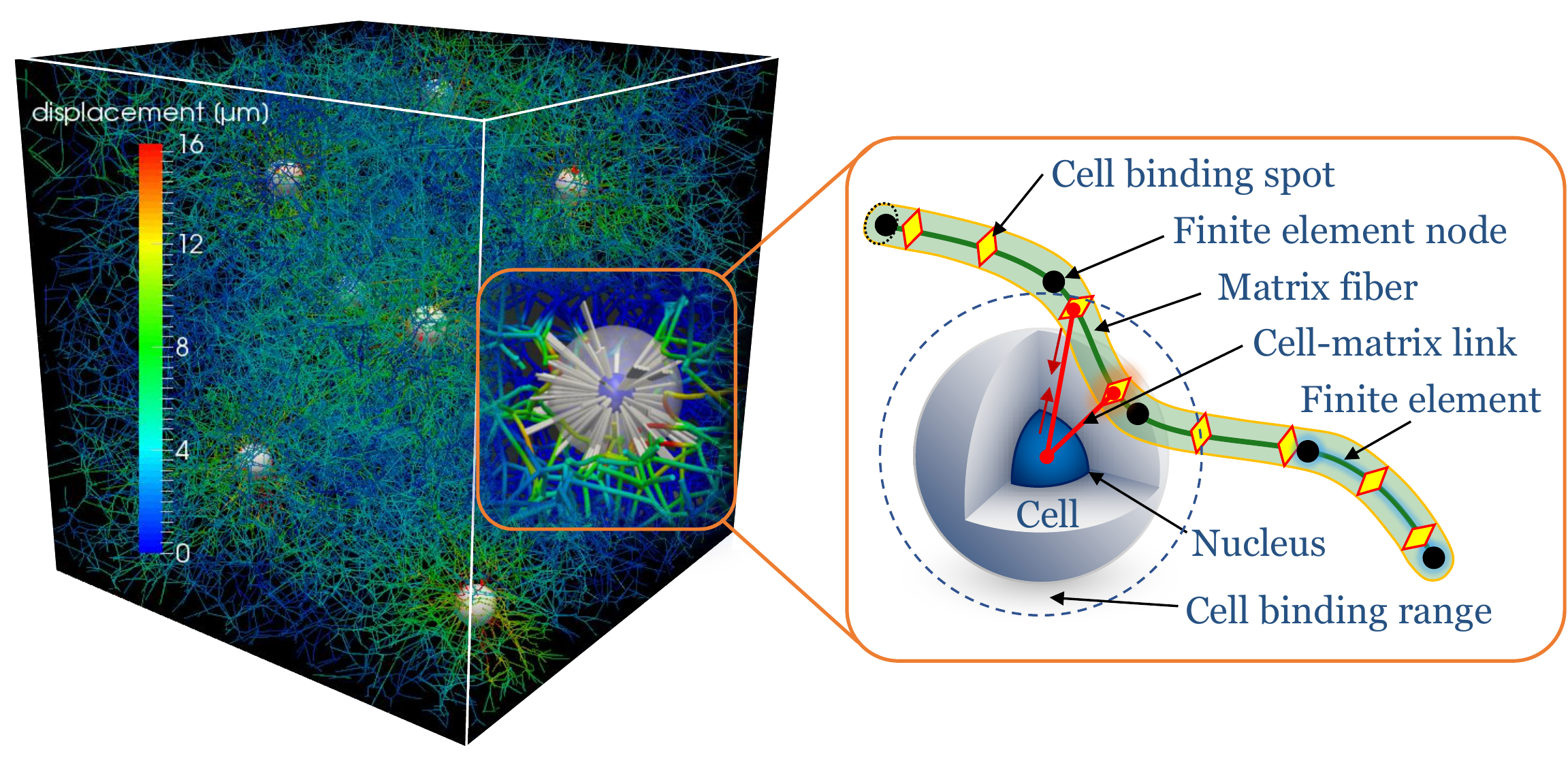}
\end{center}  
\caption{RVE of a three-dimensional discrete fiber and cell model. Fibers are modeled as nonlinear beam elements, on which cells can exert contractile forces via elastic links representing focal adhesions.}
\label{fig:methods_discrete_fiber_model}       
\end{figure}

To simulate of soft tissue mechanics on the level of individual cells and fibers, we used the computational framework presented in \cite{Eichinger2020c} (Fig. \ref{fig:methods_discrete_fiber_model}). Briefly, we constructed periodic RVEs of fiber networks that neatly matched the crucial microstructural characteristics of the actual collagen gels, that is, their valency, free fiber length, and orientation distributions. Individual fibers were discretized with nonlinear beam finite elements, which are well-known to capture the mechanical behavior of fibers. Covalent bonds between fibers were modeled as rigid joints. Fibers were assumed to have circular cross-sections with a diameter of $180 nm$ \cite{VanDerRijt2006a} and an elastic modulus of $1.1 MPa$ \cite{Jansen2018}. Cells were represented by spherical particles exerting tensile forces on nearby fibers via elastic connections. They were distributed in a manner that closely resembles the biophysical processes in focal adhesions in which integrins connect cells and matrix fibers. All simulations were performed using displacement-controlled boundary conditions for the considered RVEs. The entire computational framework was implemented in the in-house finite element code BACI \cite{Baci2020}.

\section{Results}
\label{sec:results}

\subsection{Experimental results}
\label{subsec:results:exp}

\begin{figure*}[htbp] 
\begin{minipage}[c]{0.0\linewidth}
\end{minipage}
\begin{minipage}[c]{0.0\linewidth}
\end{minipage}
\hfill
\begin{minipage}[c]{0.32\linewidth}
A
\end{minipage}
\hfill
\begin{minipage}[c]{0.32\linewidth}
B
\end{minipage}
\hfill
\begin{minipage}[c]{0.32\linewidth}
C
\end{minipage}
\hfill
\begin{minipage}[c]{0.0\linewidth}
\end{minipage}
\hfill
\\
\begin{minipage}[c]{0.0\linewidth}
\end{minipage}
\hfill
\begin{minipage}[c]{0.32\linewidth}
	\centering
	\includegraphics[width=1.0\linewidth]{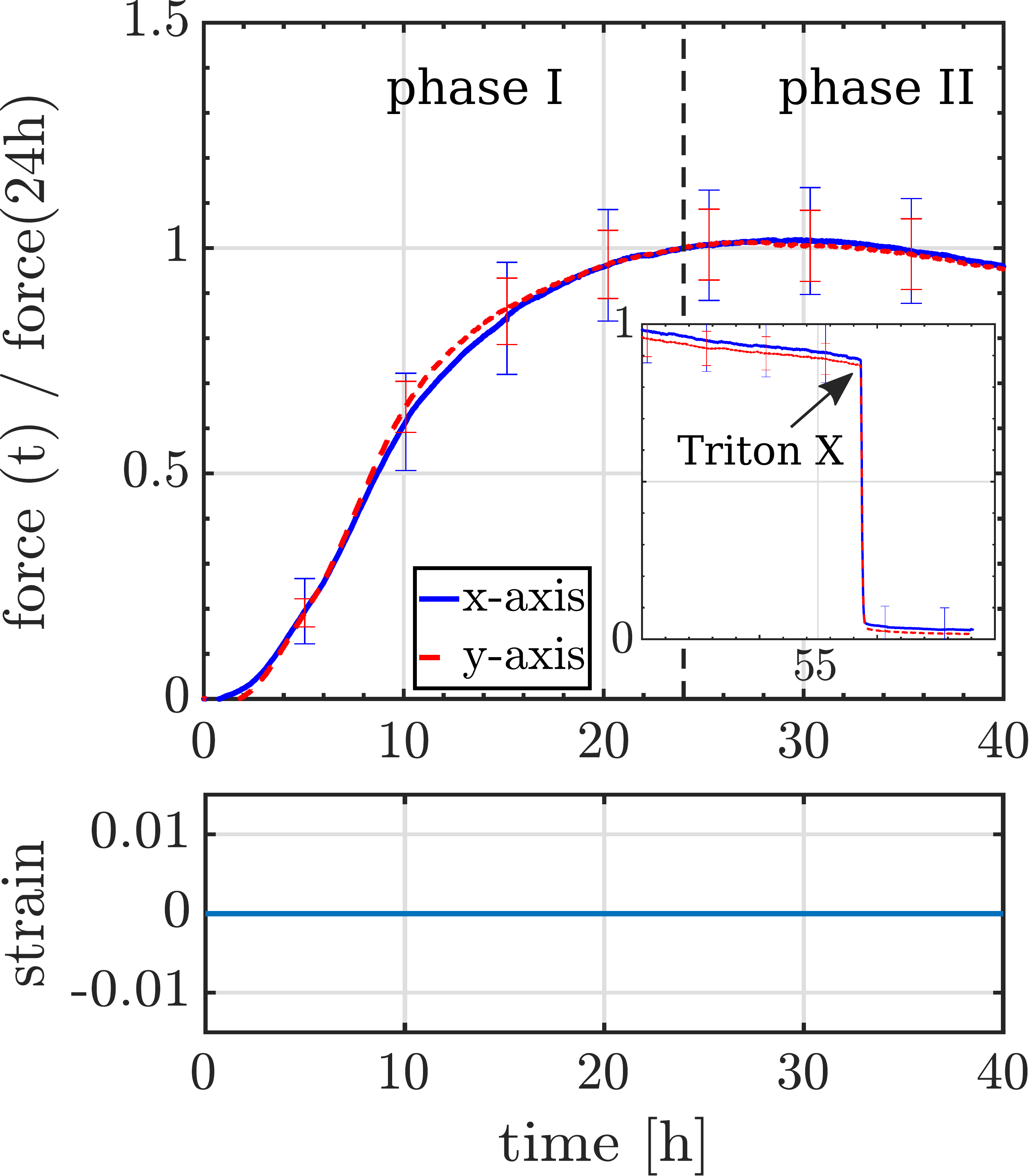}
\end{minipage}
\hfill
\begin{minipage}[c]{0.30\linewidth}
	\centering
	\includegraphics[width=1.0\linewidth]{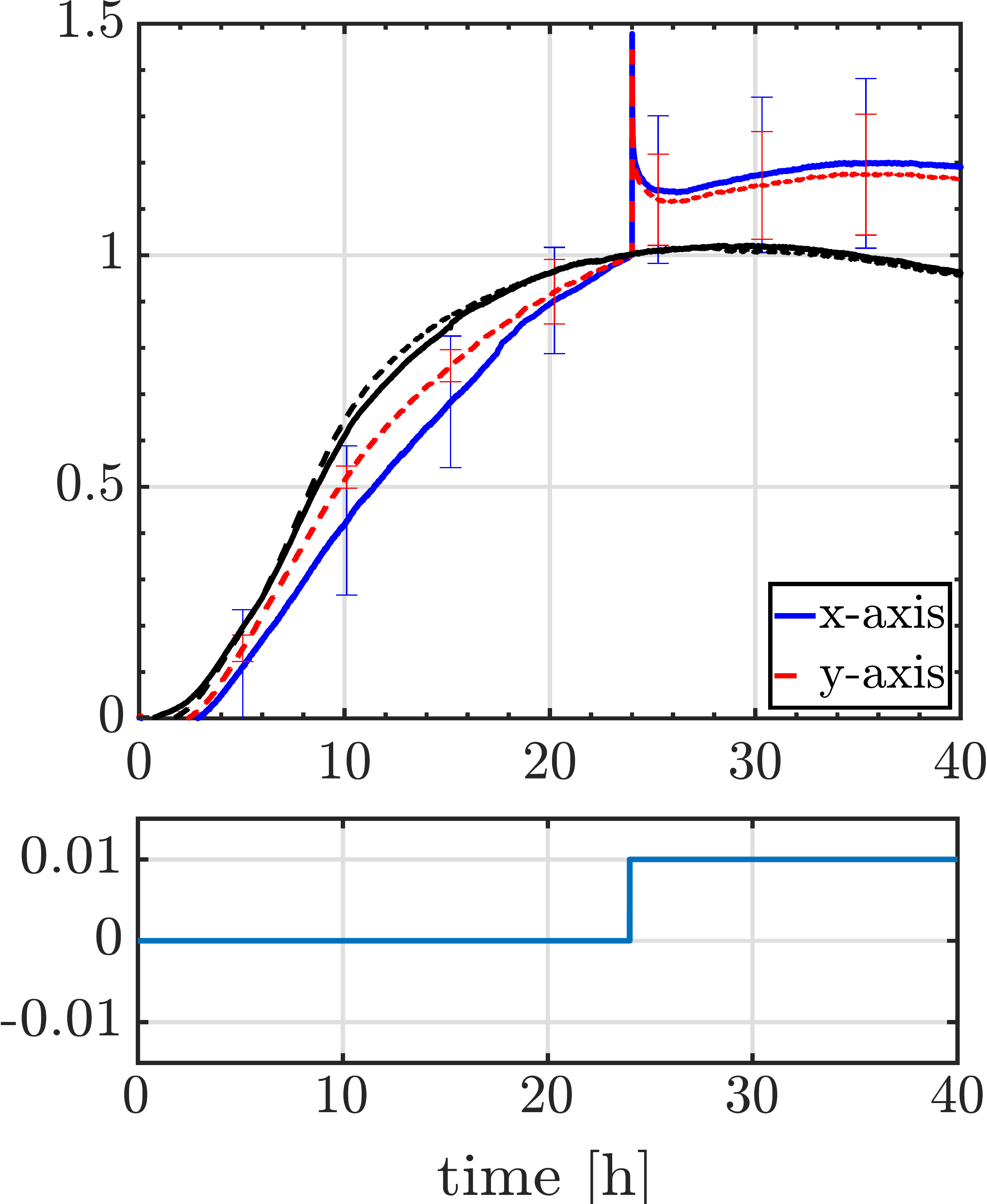}
\end{minipage}
\hfill
\begin{minipage}[c]{0.30\linewidth}
	\centering
	\includegraphics[width=1.0\linewidth]{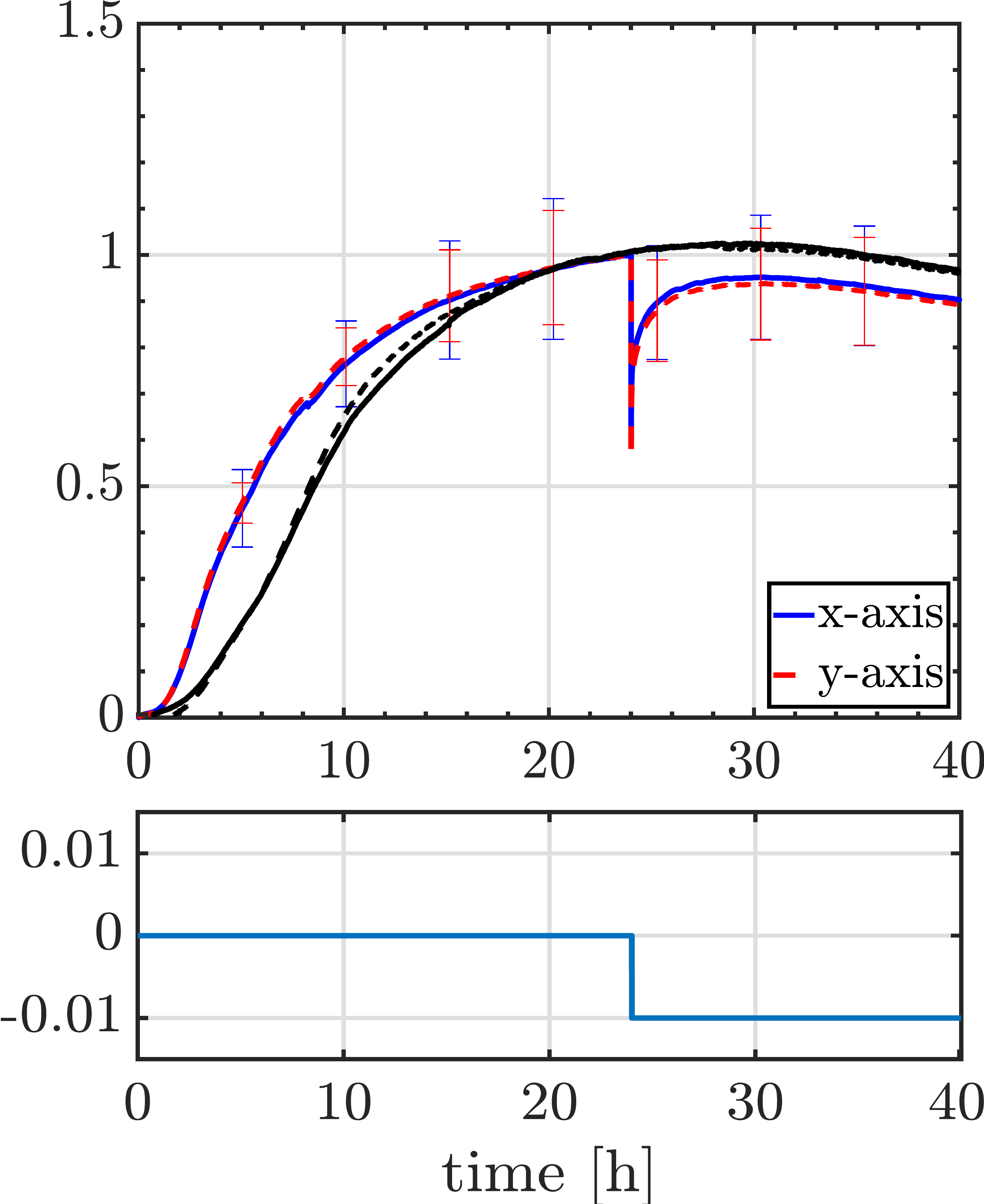}
\end{minipage}
\hfill
\begin{minipage}[c]{0.0\linewidth}
\end{minipage}
\hfill
\hfill
\begin{minipage}[c]{0.0\linewidth}
\end{minipage}
\hfill
\caption{Normalized force in cruciform-shaped collagen gels (arms of the gel aligned with x- and y-axis, respectively) seeded with primary aortic SMCs. Each curve shows the mean $\pm$ SEM of three identical experiments using a collagen concentration of $1.0mg/ml$ and a cell density of $0.5\cdot 10^6 \text{cells}/ml$. (A) Unperturbed tissue equivalents (normalized with $F_x(24h) = 720 \mu N$ and $F_y(24h) = 729 \mu N$) suggested a nearly isotropic biaxial response. (B) Tissue equivalents perturbed with a strain step of $2\%$ at $24 h$ (normalized with $F_x(24h) = 602 \mu N $ and $F_y(24h) = 588 \mu N$). (C) tissue equivalents perturbed with a step-wise release by $2\%$ at $24 h$ (normalized with $F_x(24h) = 664 \mu N$ and $F_y(24h) = 626 \mu N$). Lines without error bars in (B) and (C) represent experiments without perturbation from (A), hence revealing some specimen-to-specimen variations.}
\label{fig:exp_smc}
\end{figure*}

Experimental studies of cell-seeded collagen gels (tissue equivalents) subject to mechanical perturbations so far largely suffer from the unsatisfactory short periods over which the gels have been monitored after the perturbations (e.g. only $30 min$ in \cite{Brown1998,Ezra2010}). Therefore, it has remained largely unclear so far whether tissue equivalents recover the prior state of tension or only to some extent after perturbations. To close this gap, we performed our experiments with cruciform-shaped tissue equivalents (leading to uniaxially loaded arms and a biaxially loaded central region) over prolonged periods up to $40 h$. After $24 h$ we strained some of the gels by $2\%$ and $-2\%$, respectively, granting another $16 h$ for the observation of the resulting period of recovery. Importantly, our gels showed neither growth nor remodeling, as the addition of Triton X after 40h to induce cell lysis led to a rapid decrease of the gel tension to zero (\ref{fig:exp_smc} A inset ), implying that all forces measured were actively applied by the cells, with no appreciable inelastic matrix deformations or entrenchment (e.g. via transglutaminases). Similar results were found before \cite{Karamichos2007}.

In this setting, we initially observed the well-known increase of tension up to a homeostatic plateau \cite{Brown1998,Ezra2010,Marenzana2006,Brown2002,Courderot-Masuyer2017,Campbell2003a,Dahlmann-Noor2007,Karamichos2007,Sethi2002}. Also as previously reported for porcine SMCs \cite{Hall2007}, this first stage was followed by a slight decline of tension (Fig. \ref{fig:exp_smc} A), possibly due to some form of exhaustion of the cells. In cases where tissue equivalents were strained by a $2\%$ step at $24 h$ (leading to a step-like perturbation of $F_t$ of $\sim 50\% $), gel tension first increased in a step-wise manner (elastic response of cells and matrix) followed by a period where tension decreased back towards the level prior to the perturbation (some isolated cellular response). However, even after $16 h$, the original level of tension was not fully recovered, but rather re-established within $\sim 10-15\%$ deviation from the prior value (Fig. \ref{fig:exp_smc} B). Analogously, if the gels were released by $2\%$ at $24 h$ (leading to a step-like perturbation of $F_t$ of $\sim 40\% $), one first observed a step-wise drop of tension (elastic response of cells and matrix), followed by a period where tension increased back towards the level prior to the perturbation (some isolated cellular response). Again, however, even after $16 h$ the original level of tension was not fully recovered (Fig. \ref{fig:exp_smc} C), but rather re-established within $\sim 5-10\%$ deviation from the prior value.

\subsection{Mechanical analog model}
\label{subsec:results_analog}

The primary observation of the previous section is: when cell-seeded tissue equivalents were perturbed from the apparent homeostatic state achieved over $24h$, they did not recover precisely $F_t$ (over periods shorter than 2 days). To understand the origin of this behavior, we employed the mechanical analog model introduced in Section \ref{sec:methods}. In this model, the external force on the tissue $F_t$ needs to equal the elastic force $F_1$ in the tissue region under tension in series with the cells, which has to balance the sum of the cellular force $F_c$ and the elastic forces $F_2$ of the elastic forces in the tissue region under compression. This yields 
\begin{align}
F_t &= F_1 = F_2 + F_c.
\label{eqn:force_equilibrium}
\end{align}
We now assume the system to be in a homeostatic state (Fig. \ref{fig:explanation_for_force_offset} A), in which the initially stress-free regions 1 and 2 were deformed by tensile cell forces $F_c > 0$. One can easily show that this results in an initial homeostatic force on tissue level
\begin{align}
F_{t0} &= \left( 1 - \frac{k_2}{k_1 + k_2} \right)F_c.
\label{eqn:Ft0}
\end{align}
We then subject the tissue to a step-wise stretch or release by a change of length $\Delta L_t$ (Fig. \ref{fig:explanation_for_force_offset} B). Keeping $\Delta L_t$ constant after the perturbation results in a permanent change of tissue length
\begin{align}
\Delta L_t = \Delta L_1 + \Delta L_2,
\label{eqn:hypI:L}
\end{align}
which is composed of the accumulated fiber length changes $\Delta L_1$ and $\Delta L_2$. The elastic response of the system will be a step-wise increase of $F_t$, the quantity that can be measured externally. The subsequent evolution of the forces in the tissue is directly governed by cellular mechanoregulation if we assume that the fiber network only deforms elastically (neither growth nor inelastic remodeling on the short time scales considered). 

In the following, we discuss on the basis of our mechanical analog model competing hypotheses regarding the quantity that cells actually regulate. We discuss the observations on the macroscale that these hypotheses would yield and compare them with those in our experiments. It appears reasonable to assume that cells can sense and thus regulate on the microscale three quantities (cf. also \cite{Humphrey2001}): their own shape (hypothesis I), the active force they exert through their focal adhesions (hypothesis II), or the strain of the fibers to which they are connected by focal adhesions (hypothesis III). This yields three hypotheses which will be discussed in the following. We assume for simplicity a linear-elastic behavior, that is, deformation-independent stiffnesses $k_i$.

\subsubsection{Hypothesis I: cells restore their shape}
\label{subsubsec:hypo_shape}
\begin{figure*}[htb!] 
\begin{center}
 \includegraphics[width=0.82\textwidth]{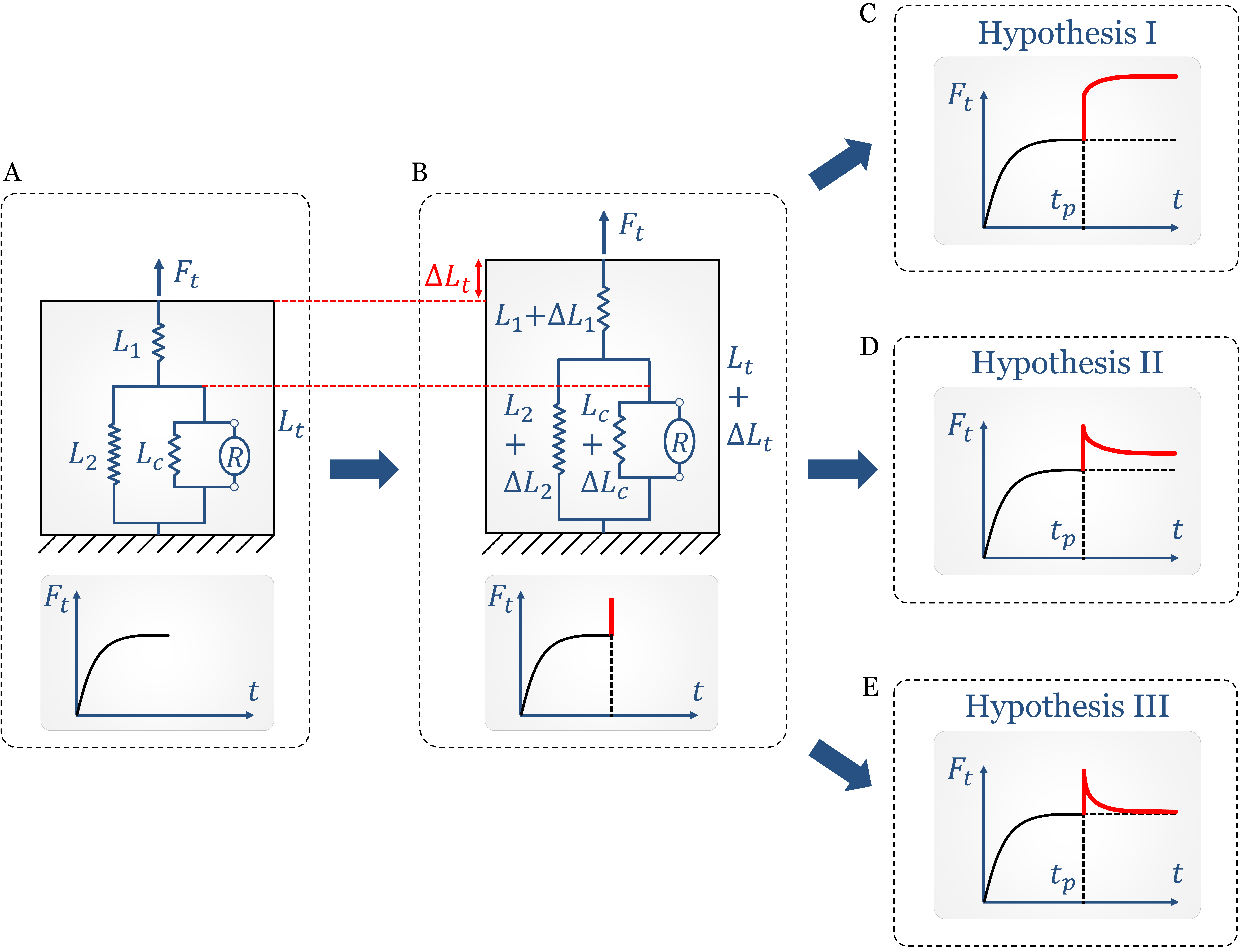}
\end{center}  
\caption{Short-term response of the (A) mechanical analog model at steady state (after the force $F_c > 0 $ built up over time) to a (B) strain step assuming different regulatory targets of an individual cell formulated in (C) hypothesis I (regulation of cell shape), (D) hypothesis II (regulation of contractile forces of cells on ECM), and (E) hypothesis III (regulation of tissue strain). Note that only hypothesis II yields an answer in agreement with experimental observations.}
\label{fig:explanation_for_force_offset}       
\end{figure*}

If cells restore their shape after perturbations, they restore $L_c$ and thus also $L_2$ and $F_2$. To this end, cells have to contract after an initial step-wise stretch of the whole tissue and thus have to increase $F_c$. After regulation, $\Delta L_1 = \Delta L_t$ and with $F_2$ fully restored,
\begin{align}
\Delta F_t = \Delta F_1 = \Delta F_c = k_1 \Delta L_1 = k_1 \Delta L_t.
\label{eqn:hypII:F}
\end{align}
Therefore, $\Delta F_t$ increases its magnitude compared to that after the perturbation, which is $|\Delta F_t| = |k_1 (\Delta L_t - \Delta L_2)|$. This behavior is illustrated in (Fig. \ref{fig:explanation_for_force_offset} C), and is in contradiction to our experimental observations.

\subsubsection{Hypothesis II: cells restore cellular forces}
\label{subsubsec:hypo_force}

As discussed previously \cite{Eichinger2020b,Kong2009,Weng2016}, cells have a tendency to build stable bonds to the ECM fibers only in a certain constant range of forces. Thus, we examine the response of our analog model if cells restore the cellular forces after perturbations, i.e., $F_c$. As \eqref{eqn:force_equilibrium} must hold also for changes of forces due to changes of lengths, we have, once $F_c$ has been restored,
 \begin{align}
k_1 \Delta L_1 = \Delta F_1 = \Delta F_2 = k_2 \Delta L_2.
\label{eqn:hypI:F}
\end{align}
Combining \eqref{eqn:force_ext} - \eqref{eqn:hypI:F} yields 
 \begin{align}
\Delta F_t = \Delta F_1 = \Delta F_2 = \frac{k_1k_2}{k_1 + k_2}\Delta L_t.
\label{eqn:hypI:F_2}
\end{align}
Thus, a restoration of $F_c$ after the perturbation necessarily implies a permanent increased value of both $F_2$ and $F_1$ and thus also of $F_t$ for  $\Delta L_t > 0$, and a permanently decreased value for $\Delta L_t < 0$ (Fig. \ref{fig:explanation_for_force_offset} D). This is the behavior observed in our experiments.

Strikingly, this may suggest that most short-term tissue equivalent experiments do not study a regulation of the mechanical state of the ECM, but rather a superposition of the passive matrix response (according to \eqref{eqn:hypI:F_2} equal to the remaining offset) and the cellular regulation of a specific contractile force, which represents the relaxation (recovery) in case of extension (release) external mechanical perturbations. 

Therefore, our results agree with the findings of \cite{Weng2016}, which showed that isolated cells restore a specific cellular tensional state. Here we predict this in a three-dimensional fibrous, multi-cellular environment compared to a single cell on a two-dimensional substrate.

Moreover, the changes represented by equation \eqref{eqn:hypI:F_2} suggest a simple additional test of hypothesis II by future experiments. By performing the experiments shown herein in the future with two or more different fiber concentrations (implying different network stiffnesses \cite{Alcaraz2011,Miroshnikova2011,Joshi2018}) and measuring the resulting residual offset $\Delta F_t$, one could check whether the latter is in agreement with \eqref{eqn:hypI:F_2}. If so, it should - ceteris paribus - increase by the same factor as the network stiffness.

\subsubsection{Hypothesis III: cells restore strain in ECM fibers}
\label{subsubsec:hypo_homeo}

If cells restore the strain in the ECM fibers on which they are pulling after the prescribed perturbations, they restore $L_1$ and thereby also $F_1$ and $F_t$. Thus, hypothesis III also contradicts our experiments, where $F_t$ is not exactly restored after perturbations. To understand the problem of hypothesis III, note that it implies $\Delta L_2 = \Delta L_t$ in the long run (that is, after a step-wise elastic deformation and the subsequent mechanoregulation by the cells). It thus implies $\Delta F_2 = k_2 \Delta L_t$. With $0 = \Delta F_1 = \Delta F_2 + \Delta F_c$, one obtains
\begin{align}
\Delta F_c &= -\Delta F_2 = - k_2 \Delta L_t.
\label{eqn:delta_F_c}
\end{align}
From this equation we see a possible reason why cells apparently do not restore the strain and thereby not exactly the tension in the fibers on which they are pulling. As apparent from \eqref{eqn:delta_F_c}, they would require information about the stiffness or forces in the region under compression. However, this would require that the cells not only sense the general stiffness of the surrounding tissue, but also specifically the extensional stiffness of the part of the ECM which they compress. Moreover, cells do not have information about $\Delta L_t$. Thus, it appears that cells do not have the information necessary to regulate the strain of the fibers on which they pull, which explains why hypothesis III seems to be in disagreement with our experiments.

\subsection{Discrete fiber network model}
\label{subsec:results:fullsim}

\begin{figure}[htbp] 
\begin{minipage}[c]{0.0\linewidth}
\end{minipage}
\begin{minipage}[c]{0.0\linewidth}
\end{minipage}
\hfill
\begin{minipage}[c]{0.48\linewidth}
A
\end{minipage}
\hfill
\begin{minipage}[c]{0.48\linewidth}
B
\end{minipage}
\hfill
\begin{minipage}[c]{0.0\linewidth}
\end{minipage}
\hfill
\\
\begin{minipage}[c]{0.0\linewidth}
\end{minipage}
\hfill
\begin{minipage}[c]{0.48\linewidth}
	\centering
	\includegraphics[width=0.85\linewidth]{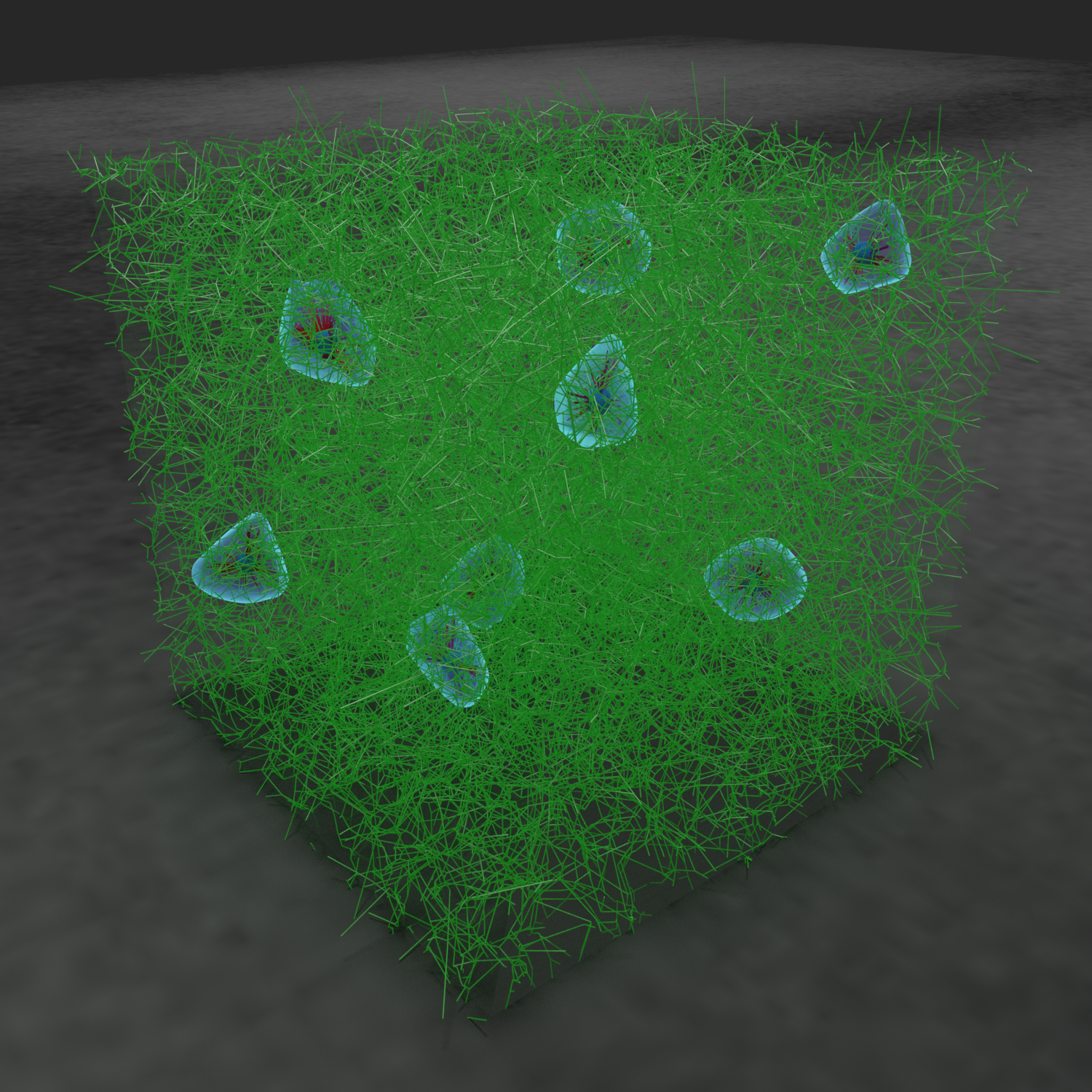}
\end{minipage}
\hfill
\begin{minipage}[c]{0.48\linewidth}
	\centering
	\includegraphics[width=0.99\linewidth]{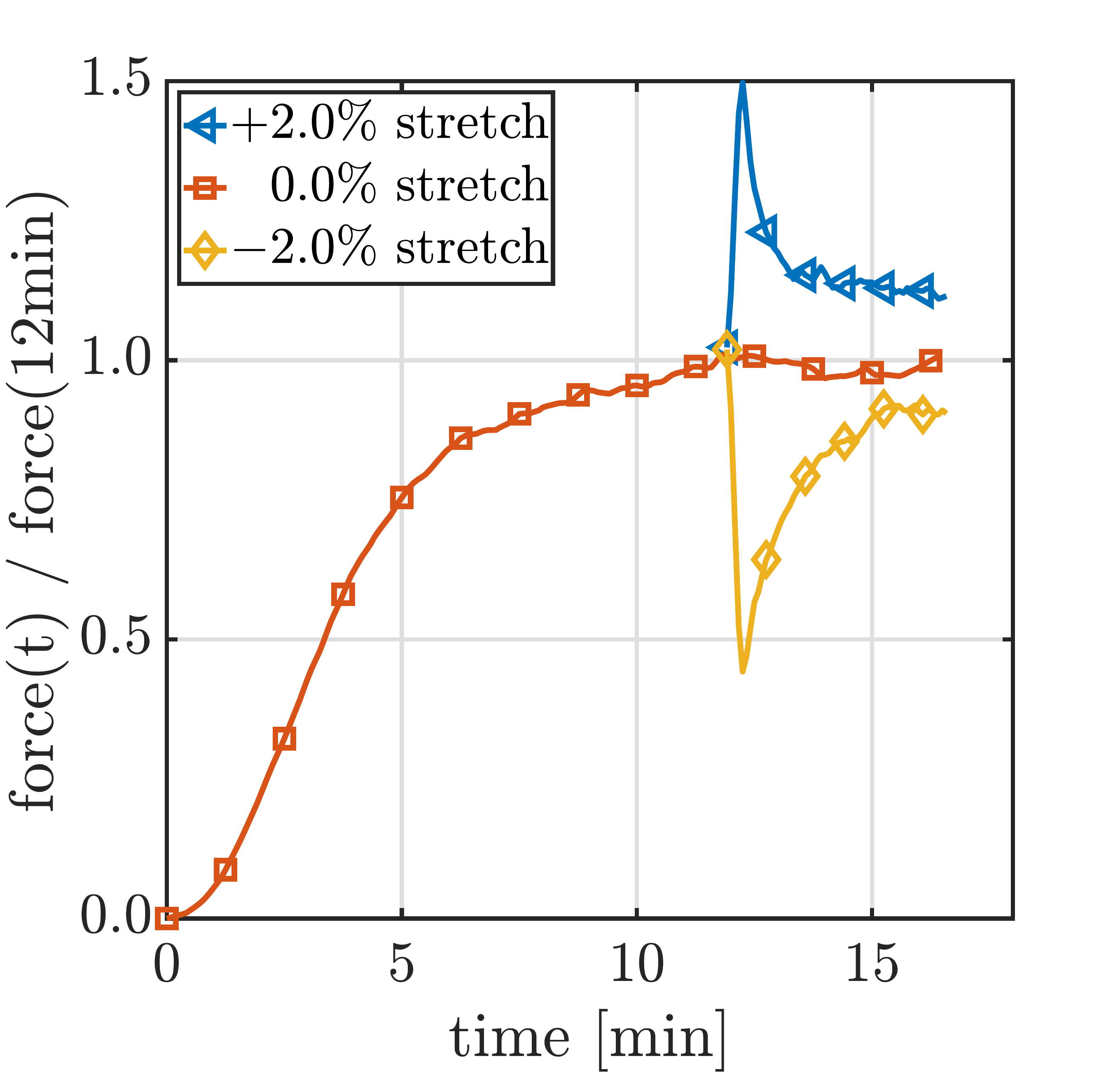}
\end{minipage}
\hfill
\begin{minipage}[c]{0.0\linewidth}
\end{minipage}
\hfill
\caption{(A) RVE simulated with a discrete fiber model; triaxial boundary conditions are applied to externally perturb the system. Cell shape is reconstructed around cell-matrix links using 3D delaunay triangulation. (B) tissue tension increases in simulations initially to a plateau value. If this plateau value is perturbed, the prior level of tension is restored towards, but not precisely to, the prior steady state value, consistent with the concept of homeostasis now with a mechanistic understanding.}
\label{fig:resulst_dfm}
\end{figure}

The main conclusion drawn from our experimental data and our simple mechanical analog model is: on short time scales, tissues do not - and in fact cannot - control their tension to a specific value. Cells only regulate the forces they exert on the surrounding fibers. This naturally leads to a residual offset in the tissue tension after perturbations on time scales too short for remodeling or de novo deposition and degradation of fibers. To corroborate this understanding of cellular mechanoregulation, we performed computer simulations with a discrete fiber-network model introduced in \cite{Eichinger2020c}. We studied an RVE with a covalently cross-linked fiber matrix (Fig. \ref{fig:resulst_dfm} A and B). The size, fiber concentration and cell concentration of the simulated RVE were chosen to be equivalent to the cell and collagen density in our experiments.

Following \cite{Kong2009}, catch-slip bonds were assumed between cells and ECM fibers. These bonds are chemically the most enduring in a very specific regime of forces. Our objective was to test whether this behavior of the catch-slip bonds together with cellular contractility alone allows cells within the overall system to effectively control the forces they exert on the surrounding fibers \cite{Weng2016} (i.e., $F_c$ in our mechanical analog model), and whether this leads to a residual offset of the tissue tension after mechanical perturbations. As Fig. \ref{fig:resulst_dfm} reveals, this is indeed the case, confirming that the catch-slip bond is a key factor enabling cells to accurately control the contractile forces they exert on surrounding ECM fibers.

\section{Discussion and conclusions}
\label{sec:discussion}

As mentioned above, the short periods of observation in previous experimental studies (e.g., \cite{Brown1998,Ezra2010}) did not allow definite conclusions with regard to which quantity is restored by mechanical homeostasis on a tissue level on short time scales following a perturbation from the homeostatic state by imposition of a step-wise stretch or release. We presented experimental data with observation periods after the perturbations that were around 30 times longer than the ones in \cite{Brown1998,Ezra2010}. This way we could observe whether the tension that develops naturally in tissue equivalents returns following a perturbation within a certain tolerance though not exactly to the prior value.

To unravel micromechanical principals underlying this behavior, we developed a mechanical analog model to test three competing hypotheses regarding what cells sense and regulate on the microscale. Hypothesis I assumed that cells regulate their own dimension. Hypothesis II, motivated by the experiments of \cite{Weng2016}, assumed that cells regulate the contractile forces they exert on the ECM. Hypothesis III assumed that cells regulate the strain in the surrounding tissue. It turned out that only Hypothesis II was consistent with the observed behavior. We therefore conclude that it is highly likely that cells in gel-like tissue equivalents (at least on short time scales) regulate only the forces they exert on the ECM rather than any tissue-intrinsic quantity.
 
Using an advanced computational model resolving discrete fibers, cells, and their interactions, we confirmed that the catch-slip bond by which integrins connect cells and matrix fibers can endow cells with an ability to regulate the contractile forces they exert on the ECM. In general, catch-slip bonds differ from most chemical bonds in that their lifetime does not monotonically decrease with increasing mechanical load on the bond. Rather there is a specific optimal loading at which the stability of these bonds attains a maximum \cite{Kong2009}. In agreement with the experiments of \cite{Weng2016}, our studies reveal that this maximum determines the level at which cells can regulate the contractile forces they exert on ECM. It is worth noting that the computational studies with our discrete fiber model can support the assumption that the catch-slip bond is sufficient for cells to regulate the forces they exert on the ECM. Yet, these studies cannot prove that this is the only mechanism by which cells can or do act in this setting.

\begin{figure}[htb!] 
\begin{center}
\includegraphics[width=0.5\textwidth]{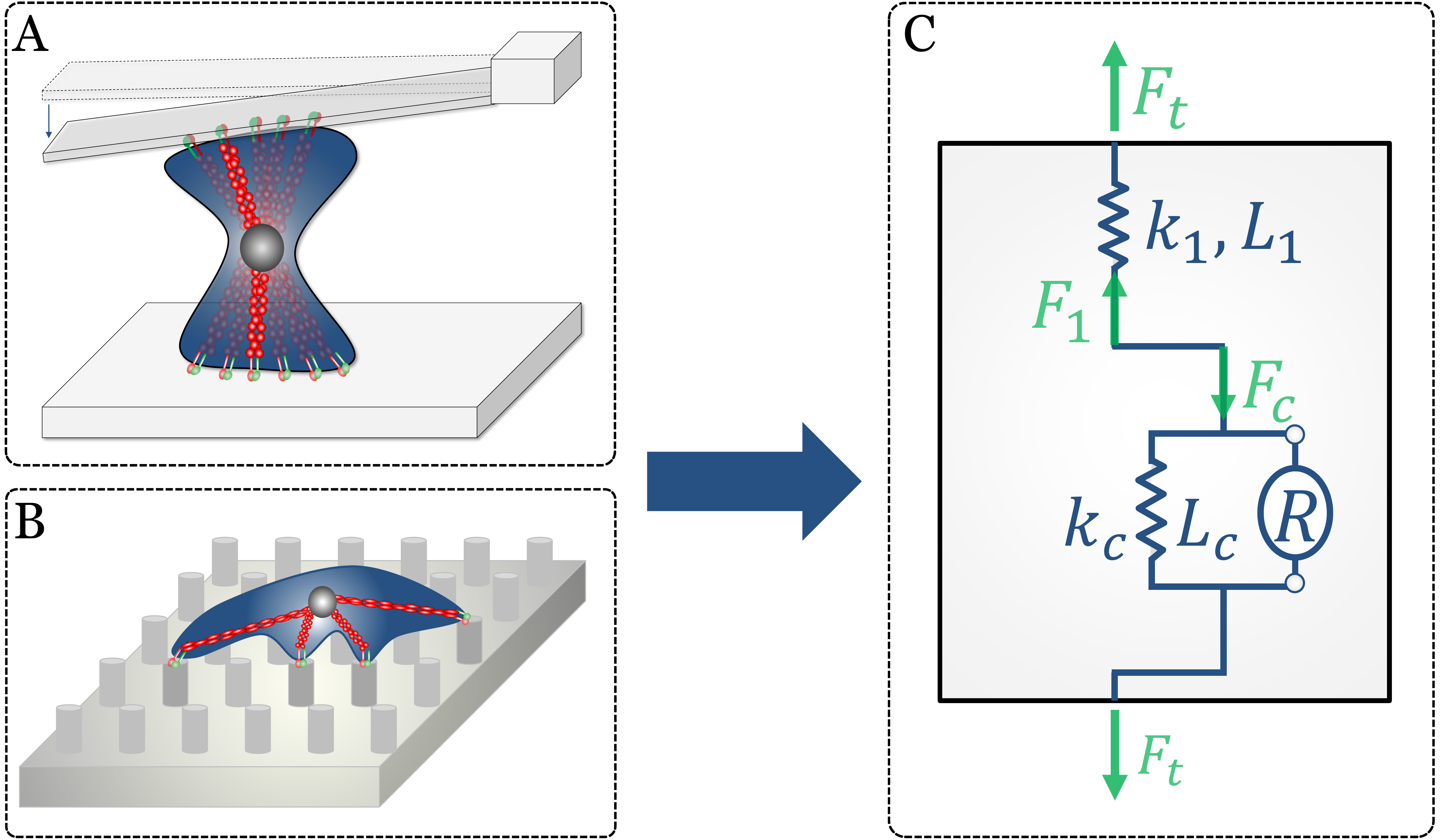}
\end{center}  
\caption{Schematic of experimental set-up used by (A) \cite{Webster2014} and (B) \cite{Weng2016} to study mechanical homeostasis on the level of single cells. Both set-ups miss the elastic fibers acting in parallel to cells in real tissues and thus an important element defining how tissues respond to perturbations of their homeostatic state.}
\label{fig:comparison_hom_param}       
\end{figure}

An important conclusion from both our mechanical analog model and simulations with our discrete fiber model is that the passive elasticity of the ECM acts in parallel to the cells to form an essential part of the mechanoregulatory system on the tissue level. Our findings suggest that the residual offset between the matrix tension before and after strain perturbations can be explained only from the passive elasticity of the ECM acting in parallel to the cells. To the authors' knowledge, this insight is new and it can be used to design future experiments. To study mechanical homeostasis on the level of single cells, \cite{Webster2014} placed cells between an elastic cantilever and a rigid substrate (Fig. \ref{fig:comparison_hom_param} A), and \cite{Weng2016} on top of a stretchable micropost array  (Fig. \ref{fig:comparison_hom_param} B). In both cases, the elastic effect of fibers acting in parallel to the contractile forces exerted by the cells is missing as illustrated in Fig. \ref{fig:comparison_hom_param} C. This means, neither of these systems mimic well that which defines mechanical homeostasis on the tissue scale. Hence, it will be essential to develop additional experimental set-ups that model this important cell-matrix interaction. 

An important question for future work is, how the conclusions drawn here can be tester further by additional experiments. As discussed above, a simple test for our conclusion, that the contractile forces exerted by cells are the quantity controlled by the cells on short time scales at a tissue level, could be performed by running the experiments shown herein with several different collagen concentrations, and observing whether the residual offset between the tissue tension before and after the perturbation scales with (approximately) the same factor as the tissue stiffness. Another way to test the conclusions of this paper would be to perform a series of experiments with a varying cell density. While the residual offset between the matrix tension before and after a perturbation was shown in \eqref{eqn:hypI:F_2} to be independent of the contractile forces of the cells in the pre-perturbed state, \eqref{eqn:Ft0} reveals that homeostatic tissue tension prior to the perturbation linearly scales with the magnitude of these forces. That is, the discussion of this paper suggests a decreasing relative offset of tension before and after the perturbation as the cell density increases. Future experiments with varying cell density can easily test this.

In summary, the central result of this paper is that, on short time scales that preclude deposition and degradation of ECM, mechanical homeostasis on the tissue level likely results primarily from the contractile forces exerted by the cells on the surrounding tissue. Cells thereby re-establish after perturbations a state only similar to the one prior to the perturbation. Using the mechanical analog model and computational framework presented in this paper to study the response of cell-seeded collagen gels and soft tissues to perturbations on longer time scales is a promising avenue of future research.

\section*{Acknowledgments}
This work was supported by the Deutsche Forschungsgemeinschaft (DFG, German Research Foundation) – Projektnummer 257981274, Projektnummer 386349077. The authors gratefully acknowledge financial support by the International Graduate School of Science and Engineering (IGSSE) of Technical University of Munich, Germany. In addition, we thank Lydia Ehmer and Lea Haeusel for conducting some of the experiments. We further gratefully thank Diane Tchibozo and Lisa Pretsch for contributing to Fig. 6A.
\end{document}